\documentclass[aps,prl,twocolumn,superscriptaddress]{revtex4-2}
\usepackage{graphicx}% Include figure files
\usepackage{makecell}
\usepackage{booktabs}
\usepackage{multirow}
\usepackage{color}
\usepackage[dvipsnames]{xcolor}
\usepackage[colorlinks=true,urlcolor=Blue,linkcolor=Blue]{hyperref}
\usepackage[all]{hypcap}

\begin{document}

\title{$B\rho$-defined isochronous mass spectrometry: a new approach for high-precision mass measurements of short-lived nuclei}% Force line breaks 

\author{M. Wang}
	\email{wangm@impcas.ac.cn}
	\affiliation{CAS Key Laboratory of High Precision Nuclear Spectroscopy,   Institute of Modern Physics, Chinese Academy of Sciences, Lanzhou 730000, China}
	\affiliation{School of Nuclear Science and Technology, University of Chinese Academy of Sciences, Beijing 100049, China}	
\author{M. Zhang}
	\affiliation{CAS Key Laboratory of High Precision Nuclear Spectroscopy,  Institute of Modern Physics, Chinese Academy of Sciences, Lanzhou 730000, China}
	\affiliation{School of Nuclear Science and Technology, University of Chinese Academy of Sciences, Beijing 100049, China}
\author{X. Zhou}
	\affiliation{CAS Key Laboratory of High Precision Nuclear Spectroscopy,  Institute of Modern Physics, Chinese Academy of Sciences, Lanzhou 730000, China}
	\affiliation{School of Nuclear Science and Technology, University of Chinese Academy of Sciences, Beijing 100049, China}
\author{Y. H. Zhang}
	\email{yhzhang@impcas.ac.cn}
	\affiliation{CAS Key Laboratory of High Precision Nuclear Spectroscopy,   Institute of Modern Physics, Chinese Academy of Sciences, Lanzhou 730000, China}
	\affiliation{School of Nuclear Science and Technology, University of Chinese Academy of Sciences, Beijing 100049, China}	
\author{Yu. A. Litvinov}
	\email{y.litvinov@gsi.de}
	\affiliation{CAS Key Laboratory of High Precision Nuclear Spectroscopy,   Institute of Modern Physics, Chinese Academy of Sciences, Lanzhou 730000, China}
	\affiliation{GSI Helmholtzzentrum f{\"u}r Schwerionenforschung, Planckstra{\ss}e 1, 64291 Darmstadt, Germany}	
\author{H. S. Xu}
	\affiliation{CAS Key Laboratory of High Precision Nuclear Spectroscopy,   Institute of Modern Physics, Chinese Academy of Sciences, Lanzhou 730000, China}
	\affiliation{School of Nuclear Science and Technology, University of Chinese Academy of Sciences, Beijing 100049, China}
\author{R. J. Chen}
	\affiliation{CAS Key Laboratory of High Precision Nuclear Spectroscopy,   Institute of Modern Physics, Chinese Academy of Sciences, Lanzhou 730000, China}
	\affiliation{GSI Helmholtzzentrum f{\"u}r Schwerionenforschung, Planckstra{\ss}e 1, 64291 Darmstadt, Germany}
\author{H. Y. Deng}
	\affiliation{CAS Key Laboratory of High Precision Nuclear Spectroscopy,   Institute of Modern Physics, Chinese Academy of Sciences, Lanzhou 730000, China}
	\affiliation{School of Nuclear Science and Technology, University of Chinese Academy of Sciences, Beijing 100049, China}
\author{C. Y. Fu}
	\affiliation{CAS Key Laboratory of High Precision Nuclear Spectroscopy,   Institute of Modern Physics, Chinese Academy of Sciences, Lanzhou 730000, China}
\author{W. W. Ge}
	\affiliation{CAS Key Laboratory of High Precision Nuclear Spectroscopy,   Institute of Modern Physics, Chinese Academy of Sciences, Lanzhou 730000, China}
\author{H. F. Li}
	\affiliation{CAS Key Laboratory of High Precision Nuclear Spectroscopy,   Institute of Modern Physics, Chinese Academy of Sciences, Lanzhou 730000, China}
	\affiliation{School of Nuclear Science and Technology, University of Chinese Academy of Sciences, Beijing 100049, China}
\author{T. Liao}
	\affiliation{CAS Key Laboratory of High Precision Nuclear Spectroscopy,   Institute of Modern Physics, Chinese Academy of Sciences, Lanzhou 730000, China}
	\affiliation{School of Nuclear Science and Technology, University of Chinese Academy of Sciences, Beijing 100049, China}
\author{S. A. Litvinov}
	\affiliation{GSI Helmholtzzentrum f{\"u}r Schwerionenforschung, Planckstra{\ss}e 1, 64291 Darmstadt, Germany}
	\affiliation{CAS Key Laboratory of High Precision Nuclear Spectroscopy,   Institute of Modern Physics, Chinese Academy of Sciences, Lanzhou 730000, China}
\author{P. Shuai}
	\affiliation{CAS Key Laboratory of High Precision Nuclear Spectroscopy,   Institute of Modern Physics, Chinese Academy of Sciences, Lanzhou 730000, China}
\author{J. Y. Shi}
	\affiliation{CAS Key Laboratory of High Precision Nuclear Spectroscopy,   Institute of Modern Physics, Chinese Academy of Sciences, Lanzhou 730000, China}
	\affiliation{School of Nuclear Science and Technology, University of Chinese Academy of Sciences, Beijing 100049, China}
\author{M. Si}
	\affiliation{CAS Key Laboratory of High Precision Nuclear Spectroscopy,   Institute of Modern Physics, Chinese Academy of Sciences, Lanzhou 730000, China}
	\affiliation{School of Nuclear Science and Technology, University of Chinese Academy of Sciences, Beijing 100049, China}
\author{R. S. Sidhu}
	\affiliation{GSI Helmholtzzentrum f{\"u}r Schwerionenforschung, Planckstra{\ss}e 1, 64291 Darmstadt, Germany}
\author{Y. N. Song}
	\affiliation{CAS Key Laboratory of High Precision Nuclear Spectroscopy,   Institute of Modern Physics, Chinese Academy of Sciences, Lanzhou 730000, China}
	\affiliation{School of Nuclear Science and Technology, University of Chinese Academy of Sciences, Beijing 100049, China}
\author{M. Z. Sun}
	\affiliation{CAS Key Laboratory of High Precision Nuclear Spectroscopy,   Institute of Modern Physics, Chinese Academy of Sciences, Lanzhou 730000, China}
\author{S. Suzuki}
	\affiliation{CAS Key Laboratory of High Precision Nuclear Spectroscopy,   Institute of Modern Physics, Chinese Academy of Sciences, Lanzhou 730000, China}
\author{Q. Wang}
	\affiliation{CAS Key Laboratory of High Precision Nuclear Spectroscopy,   Institute of Modern Physics, Chinese Academy of Sciences, Lanzhou 730000, China}
	\affiliation{School of Nuclear Science and Technology, University of Chinese Academy of Sciences, Beijing 100049, China}
\author{Y. M. Xing}
	\affiliation{CAS Key Laboratory of High Precision Nuclear Spectroscopy,   Institute of Modern Physics, Chinese Academy of Sciences, Lanzhou 730000, China}
\author{X. Xu}
	\affiliation{CAS Key Laboratory of High Precision Nuclear Spectroscopy,   Institute of Modern Physics, Chinese Academy of Sciences, Lanzhou 730000, China}
\author{T. Yamaguchi}
	\affiliation{Department of Physics, Saitama University, Saitama 338-8570, Japan}
\author{X. L. Yan}
	\affiliation{CAS Key Laboratory of High Precision Nuclear Spectroscopy,   Institute of Modern Physics, Chinese Academy of Sciences, Lanzhou 730000, China}
\author{J. C. Yang}
	\affiliation{CAS Key Laboratory of High Precision Nuclear Spectroscopy,   Institute of Modern Physics, Chinese Academy of Sciences, Lanzhou 730000, China}
	\affiliation{School of Nuclear Science and Technology, University of Chinese Academy of Sciences, Beijing 100049, China}
\author{Y. J. Yuan}
	\affiliation{CAS Key Laboratory of High Precision Nuclear Spectroscopy,   Institute of Modern Physics, Chinese Academy of Sciences, Lanzhou 730000, China}
	\affiliation{School of Nuclear Science and Technology, University of Chinese Academy of Sciences, Beijing 100049, China}
\author{Q. Zeng}
	\affiliation{School of Nuclear Science and Engineering, East China University of Technology, Nanchang 330013, China}
\author{X. H. Zhou}
	\affiliation{CAS Key Laboratory of High Precision Nuclear Spectroscopy,   Institute of Modern Physics, Chinese Academy of Sciences, Lanzhou 730000, China}
	\affiliation{School of Nuclear Science and Technology, University of Chinese Academy of Sciences, Beijing 100049, China}

\date{\today}

\begin{abstract}
A novel technique for broadband high-precision mass measurements of short-lived exotic nuclides is reported.
It is based on the isochronous mass spectrometry (IMS) and 
%meanwhile  
realizes simultaneous determinations of revolution time and velocity of short-lived stored ions at the cooler storage ring CSRe in Lanzhou.
The new technique, named as the $B\rho$-defined IMS or $B\rho$-IMS, boosts the efficiency, sensitivity, and accuracy of mass measurements, and is applied here to measure masses of neutron-deficient $fp$-shell nuclides. In a single accelerator setting, masses of $^{46}$Cr, $^{50}$Fe and $^{54}$Ni are determined
with relative uncertainties of (5~-~6)$\times10^{-8}$, thereby improving the input data for testing the unitarity of the Cabibbo-Kobayashi-Maskawa quark mixing matrix. This is the technique of choice for future high-precision measurements of the most rarely produced shortest-lived nuclides.
\end{abstract}

\maketitle

The mass or equivalently the binding energy of an atomic nucleus is a fundamental property
which reflects all the interactions between
%acting among 
constituent nucleons.
Thanks to more than one century of efforts, masses of about 2550 nuclides~\cite{Huang2021} have been measured,
greatly advancing our knowledge of nuclear structure and astrophysics, as well as of fundamental interactions and symmetries~\cite{Lunney2003,Blaum2006,ERONEN2016259,Dilling2018,YAMAGUCHI2021}.
The challenge today is to determine--preferably with high precision--masses of exotic nuclei with very short lifetimes and tiny production rates,
which are 
needed for the investigation of
%highly demanded to investigate 
a plethora of 
%fine 
basic
physics problems~\cite{ERONEN2016259,Dilling2018,YAMAGUCHI2021}.
For example, studies of superallowed $0^+ \to 0^+$ nuclear $\beta$ decays~\cite{Hardy2020} 
are decisive for testing the unitarity of the Cabibbo-Kobayashi-Maskawa quark mixing matrix,
which requires 
%ever improving 
precision 
%of the 
$\beta$ decay energies deduced as the mass differences of the corresponding nuclides. 
In the latest survey~\cite{Hardy2020}, the majority of decay energies were obtained from Penning trap (PT) measurements, while the heaviest $T_z=(N-Z)/2=-1$ superallowed $0^+\rightarrow0^+$ $\beta$-emitters $^{46}$Cr, $^{50}$Fe and $^{54}$Ni 
could as-yet be only measured by using the simplest form of isochronous mass spectrometry (IMS) in a heavy-ion storage ring~\cite{ZHANG2017}. 

Measurement techniques capable of obtaining 
%precisely the mass 
the precise mass
from a single, inevitably short-lived, particle are required.
PT mass spectrometry is widely considered to deliver the most precise nuclear masses.
Indeed, PTs produced a wealth of important results~\cite{ERONEN2016259,Dilling2018,YAMAGUCHI2021}.
However, in its application, a certain restricting threshold exists concerning the half-lives and/or production rates of the investigated nuclides.
In recent years, the Multi-Reflection Time-of-Flight spectrometers (MR-TOF) have emerged as powerful instruments to study nuclei
with short half-lives and low production rates~\cite{Wienholtz2013,Mougeot2021,Beck2021}.
Nevertheless, in addition to the measurement time itself, applying these devices requires preparation steps,
like cooling and bunching of low-energy radioactive species \cite{YAMAGUCHI2021}.
At high energies, mass spectrometers coupled directly to in-flight separators need essentially no time for ion preparation.
Here, Magnetic-Rigidity Time-of-Flight ($B\rho$-TOF) mass spectrometry, implemented at radioactive ion beamlines,
has produced masses of nuclides furthest away from the stability valley~\cite{Meisel2020,Michimasa2020},
albeit with modest precision.
However, the isochronous mass spectrometry (IMS)~\cite{Hausmann2000,STADLMANN2004}, based on heavy-ion storage rings~\cite{Zhang2016,STECK2020},
can achieve a much higher precision and sensitivity~\cite{Zhang2018,Xu2016}.

In the conventional IMS, a time-of-flight (TOF) detector is employed to measure revolution times of the stored ions.
A direct relation between $m/q$ and $T$ is established by using the nuclides with well-known masses as calibrants,
under the assumption that the revolution times of the stored ions are independent of their velocity or $B\rho$ spreads.
This independence is realized by the isochronous ion-optical setting of the ring requiring that the Lorentz factor $\gamma$ of stored ions is equal to the transition energy of the ring $\gamma_t$~\cite{Hausmann2000}.
A major deficiency of the conventional IMS is that the high resolving power can only be achieved in a limited range of $m/q$ values, which is termed the {\it good isochronicity region} or {\it isochronicity window} \cite{Franzke-2008}.
Due to a fixed $B\rho$ acceptance of the ring, the majority of stored ion species have $\gamma\neq\gamma_t$, and the mass resolutions of nuclides with larger or smaller $m/q$ values deteriorate rapidly.

In this Letter we present a new technique, termed the {\it $B\rho$-defined IMS} or {\it $B\rho$-IMS}, for broadband high-precision mass measurements.
It is characterized by the velocity measurements of stored ions in addition to the revolution times~\cite{Geissel2005,Geissel2006}.
In a limiting case of just a single event, the mass-to-charge ratio, $m/q$, can be determined with unprecedented precision of about 5~keV.
Therefore, this new technique is optimal for future high-precision mass measurements of the rarest short-lived nuclides.
The power of the $B\rho$-IMS is demonstrated here through the mass measurements of a series of neutron-deficient fp-shell nuclides, and the masses of $^{46}$Cr, $^{50}$Fe and $^{54}$Ni are reported with further improved precision.

\begin{figure}[htb]
	\centering
	\includegraphics[scale=1.0]{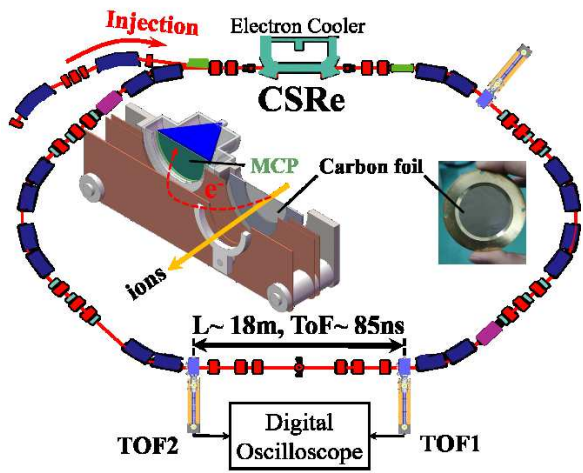}
	\caption{Schematic view of CSRe with the arrangement of two TOF detectors.}
	\label{fig:csre}
\end{figure}

The experiment was performed at the accelerator complex at the Institute of Modern Physics (IMP) in Lanzhou, China.
The nuclides of interest were produced by fragmenting the 440 MeV/u $^{58}$Ni$^{19+}$ primary beam with intensity of $8\times10^7$ particles/spill on 15~mm thick $^9$Be target. They were selected with the in-flight fragment separator RIBLL2 (second Radiactive Isotope Beam Line in Lanzhou)~\cite{XIA2002,ZHAN2010}.
Every 25 seconds, a cocktail beam including the nuclides of interest was injected into and stored in CSRe (experimental Cooler Storage Ring).
Figure~\ref{fig:csre} presents the schematic view of CSRe, which was tuned into the isochronous mode with $\gamma_t=1.365$~\cite{Zhang2022}.
The RIBLL2-CSRe system was set to a fixed central magnetic rigidity of $B\rho=5.471$~Tm. The whole $B\rho$-acceptance is about $\pm0.2\%$.
At the employed relativistic energies, the produced fragments were fully-stripped of bound electrons.

Stored ions with identical $B\rho$ values move on the same mean orbits in the storage ring.
For an ion with mass-to-charge ratio $m/q$ and velocity $v$, one can write  
\begin{equation}\label{eq:mvq}
  \frac{m}{q}=B\rho \sqrt{\Bigg(\frac{1}{v}\Bigg)^2-\Bigg(\frac{1}{v_c}\Bigg)^2}=B\rho \sqrt{\Bigg(\frac{T}{C}\Bigg)^2-\Bigg(\frac{1}{v_c}\Bigg)^2},
\end{equation}
where $C$ is the orbit length, $T$ the revolution time, and $v_c$ the speed of light in vacuum.
According to Eq.~(\ref{eq:mvq}), $m/q$ values can be determined if the quantities $(B\rho,v)$ or equivalently $(B\rho,T,C)$ are measured.

Two identical TOF detectors were installed 18~m apart in one of the straight sections of CSRe~\cite{Xing2015}.
Each detector consists of a thin carbon foil ($\phi$40~mm, 18~$\rm\mu g/cm^2$ thick) and a set of micro-channel plates (MCP)~\cite{ZHANG20141}.
When an ion passed through the carbon foil, secondary electrons were released from the foil surface and guided to MCP.
Fast timing signals from the two MCPs were recorded by an oscilloscope at a sampling rate of 50~GHz.
The duration of each measurement was merely 400~$\rm{\mu s}$.

For each ion circulating in the ring, two time sequences
were extracted from the recorded signals~\cite{Zhou2021}.
The ions stored for more than 230~$\rm{\mu s}$ were used in data analysis.
The revolution times and velocities of stored ions were deduced simultaneously following the procedures described in Refs.~\cite{XING2019,Zhou2021}.
The relative precision of velocities is at the level of $(2.2-7.2)$ $\times10^{-5}$~\cite{Zhou2021}.

On average, $\sim$15 ions were stored in one injection. Among them were also nuclides with well-known masses~\cite{Wang2021}, which were used for calibration.
Unambiguous particle identification in the measured revolution time spectrum was made~\cite{XING2019,CHEN2017}.
For each nuclide ($i$) with a well-known mass, its $(B\rho)^i_{\rm {exp}}$ and $C^i_{\rm {exp}}$ values
were deduced by using the measured $T^i_{\rm exp}$ and $v^i_{\rm exp}$ values.
The correlated dataset for $i$ nuclides, $\left\{(B\rho)_{\rm{exp}}^i,C_{\rm{exp}}^i\right\}$, was used to obtain the $B\rho(C)$ function which
%This relation 
characterizes the motion of all ions in this particular optical setting of the ring.
The magnetic fields of CSRe were not perfectly constant during the experiment, leading to an up-down shifts of the $B\rho(C)$ curve. To correct for the magnetic-field drift effects, the $B\rho$ value of every ion in each individual injection was scaled to the value corresponding to a reference field, with the scaling factor determined from experimental $B\rho$ values in this injection.
All ions with well-known masses (mass uncertainties smaller than 5~keV) and with more than 100 recorded events were used to correct for field drift effects and construct the $B\rho(C)$ function. Once the $B\rho (C)$ function is established, which is a universal calibration curve for mass determination, all $m/q$ values of stored ions, including the ions of interest, were obtained straightforwardly via Eq.~(\ref{eq:mvq}). Readers are referred to Ref.~\cite{Zhangm2022} for more technical details.

\begin{figure*}[htb]
	\centering
	\includegraphics[width=\textwidth]{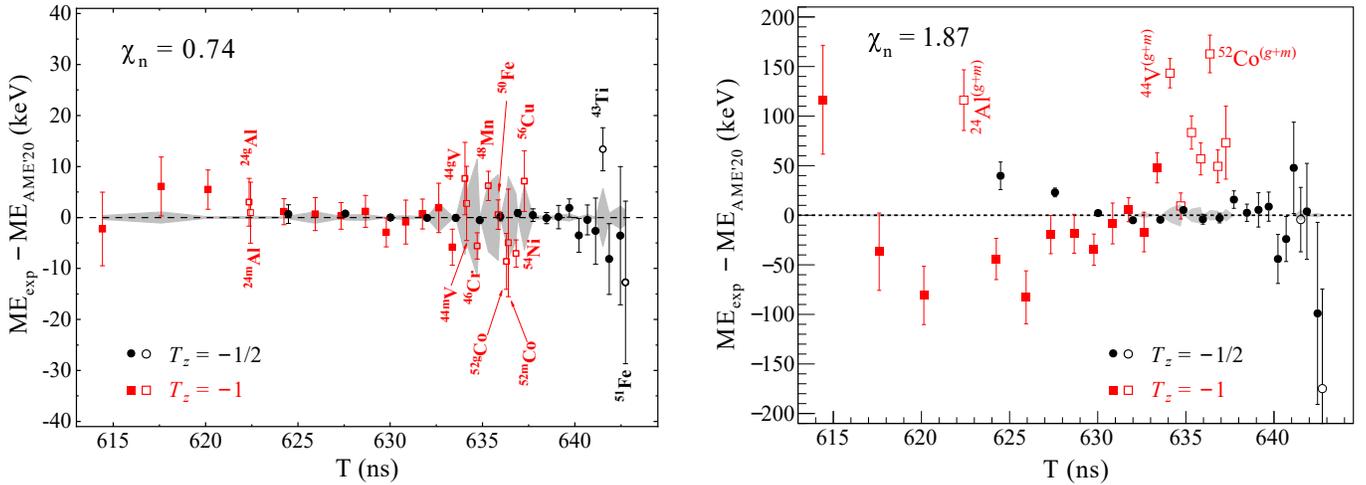}
	\caption{Left: Comparison of the re-determined masses with literature values, where the black symbols are reference nuclides and the red ones are nuclides of interest. The grey shadow represents the mass uncertainties in the AME2020~\cite{Wang2021}. Right: Comparison of the newly determined masses with the literature ones (see legend).
	}
	\label{fig:dmeresult}
\end{figure*}

%\yuhu{Although the reference nuclides} were used to derive the $B\rho(C)$ function, 
The newly determined masses are compared with literature values in Fig.~\ref{fig:dmeresult}.
The black filled squares represent the nuclides that were used to derive the $B\rho(C)$ function.
Although they are the references in the mass determination, 
their $m/q$ values can be re-determined. For this purpose, each of them was assumed to be unknown and was calculated from the remaining reference masses. A normalized $\chi_n=0.74$ for the re-determined reference masses indicates that
the quoted errors are conservative and no additional systematic errors are needed. 

%\begin{figure*}[htb]
%	\centering
%	\includegraphics[scale=0.4]{fig02_s.eps}
	%	\includegraphics[scale=0.4]{fig02_2.eps}
%	\caption{\yuhu{Comparison of the re-determined masses (red open squares) of $T_z=-1$ nuclei with literature values using the $T_z=-1/2$ nuclides as references (filled circles). (a): using conventional IMS, (b): using the $B\rho$-defined IMS. The grey shadow represents the mass uncertainties in the AME2020~\cite{Wang2021}}.
%	}
%	\label{fig:dmeresult}
%\end{figure*}

%Masses of a series of neutron-deficient nuclides were determined and are listed in Table~\ref{tab:MEvalue}.

\begin{table*}
	\caption{The numbers of identified ions $N$ and mass excess, ME, values obtained in this work, from an earlier CSRe experiment~\cite{Zhang2018}. Recent PT results for $^{44g,m}$V~\cite{Puentes2020}, $^{52g,m}$Co~\cite{Nesterenko2017}, $^{56}$Cu~\cite{Valverde2018}, $^{51}$Fe~\cite{Ong2018} and AME2016 for $^{43}$Ti~\cite{Huang2017} are included as well.
	}\label{tab:MEvalue}
	\begin{ruledtabular}
		\begin{tabular}{cccccccc}
			  &  \multicolumn{2}{c}{This work}  & \multicolumn{2}{c}{Earlier CSRe}  &  	&   &  \\\cmidrule{2-3}\cmidrule{4-5}
			Atom  &      N         & ME (keV)  &  N & ME (keV)  & $\Delta$ME (keV) 	& \makecell[c]{ME (keV)\\Literature}  & $\Delta$ME (keV) \\
			\hline
			$^{44g}$V  &  \multirow{2}{*}{601} &  $-23800.4(7.1)$& 64 & $-23827(20)$	& $-26(21)$	& $-23804.9(8.0)$~\cite{Puentes2020}	&  $-4.5(11)$   \\
			$^{44m}$V  &      & $-23534.3(7.3)$	& 75 & $-23541(19)$	& $-6(20)$	& $-23537(5.5)$~\cite{Puentes2020}	    &  $-2.7(9.1)$\\
			$^{46}$Cr  &  745 & $-29477.2(2.6)$	& 195 & $-29471(11)$	& $6(11)$	&                       &  \\
			$^{48}$Mn  &  685 & $-29290.4(2.9)$	& 198 & $-29299(7)$	& $-9(8)$	&                       &  \\
			$^{50}$Fe  &  782 &  $-34475.8(2.9)$& 342 & $-34477(6)$	& $-1(7)$	&                       &  \\
			$^{52g}$Co &  \multirow{2}{*}{845} &  $-34352.6(5.5)$& 194 & $-34361(8)$	& $-8(10)$	& $-34331.6(6.6)$~\cite{Nesterenko2017}	&  $21(9)$\\
			$^{52m}$Co &      & $-33973.0(10.6)$& 129 & $-33974(10)$	& $-2(15)$	& $-33958(11)$~\cite{Nesterenko2017}	    &  $15(15)$\\
			$^{54}$Ni  &  1254&  $-39285.4(2.7)$& 688 & $-39278.3(4.0)$& $7(5)	$	&                       &  \\
			$^{56}$Cu  &  294 &  $-38622.6(6.0)$& 64 & $-38643(15)$	& $-21(16)$	& $-38626.7(7.1)$~\cite{Valverde2018}	&  $-3.9(9.3)$\\
			$^{43}$Ti  &  757 &  $-29302.2(4.2)$& 920 & $-29306(9)$	& $-4(10)$	& $-29321(7)$~\cite{Huang2017}	    &  $19(8)$\\
			$^{51}$Fe  &  108 &  $-40201.9(15.9)$&760 & $-40198(14)$	& $4(21)$	& $-40189.2(1.4)$~\cite{Ong2018}	&  $13(16)$\\
		\end{tabular}
	\end{ruledtabular}
\end{table*}

Masses of a series of neutron-deficient nuclides were determined and are listed in Table~\ref{tab:MEvalue}.
Since the previous masses obtained by using the conventional IMS at CSRe~\cite{Zhang2018} have already been included into AME2016~\cite{Huang2017} and AME2020~\cite{Huang2021},
the present results are compared directly to older CSRe results and to the available recent PT data for $^{44g,m}$V~\cite{Puentes2020}, $^{52g,m}$Co~\cite{Nesterenko2017}, $^{56}$Cu~\cite{Valverde2018} and $^{51}$Fe~\cite{Ong2018}.

The results from this work are in excellent agreement with our earlier results~\cite{Zhang2018},
see Table~\ref{tab:MEvalue}.
% and Fig.~\ref{fig:dmeresult}.
The reliability of our previous CSRe results is mainly owing to the restriction of the $B\rho$ acceptance.
The absolute mass precision achieved here ranges from 2.6 to 16 keV thus improving most of the previous results.
The gain in mass precision is mainly due to the higher mass resolving power in a wide $m/q$ range.
For example, the present mass precision of $^{51}$Fe is nearly the same as that obtained in the previous work~\cite{Zhang2018} although this nuclide is 6 ns further away from the isochronous window and has 7 times lower statistics in this work.

%or example, the mass of $^{51}$Fe has a similar precision in this work although the statistics is 7 times smaller than that in the previous work~\cite{Zhang2018}.

\begin{figure*}[htb]
	\centering
	\includegraphics[scale=0.72]{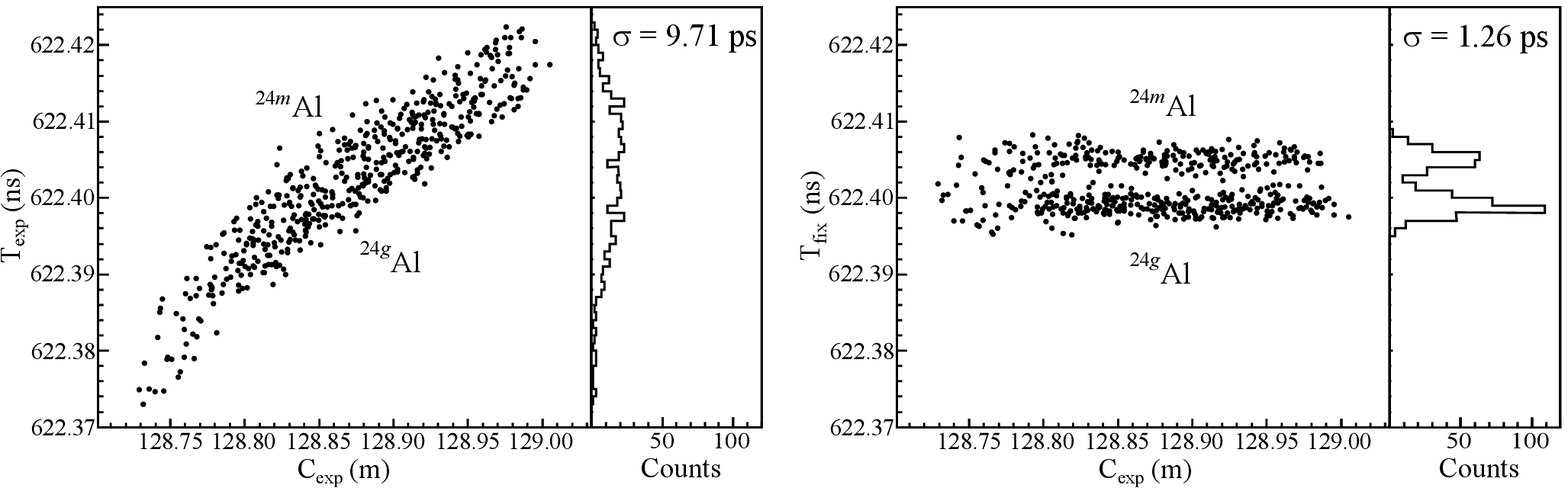}
	\caption{Scatter plots of $T_{\rm exp}$ and $T_{\rm fix}$ versus $C_{\rm exp}$ for $^{24}$Al$^{13+}$ ions (see text for details). 
	}
	\label{fig:t24al}
\end{figure*}

To demonstrate the power of the $B\rho$-IMS as comparing to the conventional IMS,
we transformed the $m/q$ spectrum into a new revolution time spectrum, $T_{\rm fix}$, at a fixed magnetic rigidity, $(B\rho)_{\rm fix}=5.4758$ Tm, and a fixed orbit length, $C_{\rm fix}=128.86$ m, according to
\begin{equation}\label{eq:tfix}
    T_{\rm fix} = C_{\rm fix}\sqrt{\frac{1}{(B\rho)_{\rm fix}^2}\left(\frac{m}{q}\right)^2 + \left(\frac{1}{v_c}\right)^2}.
\end{equation}
For example, scatter plots of $T_{\rm exp}$ and $T_{\rm fix}$ versus $C_{\rm exp}$ for $^{24}{\rm Al}^{13+}$ ions are shown in Fig.~\ref{fig:t24al}.
It is obvious that the two states in $^{24}$Al, separated by the mass difference of 425.8(1)~keV~\cite{Kondev2021}, cannot be resolved in the $T_{\rm exp}$ spectrum,
while the two peaks can clearly be separated in the $T_{\rm fix}$ spectrum.

\begin{figure}[htb]
	\centering
	\includegraphics[scale=0.4]{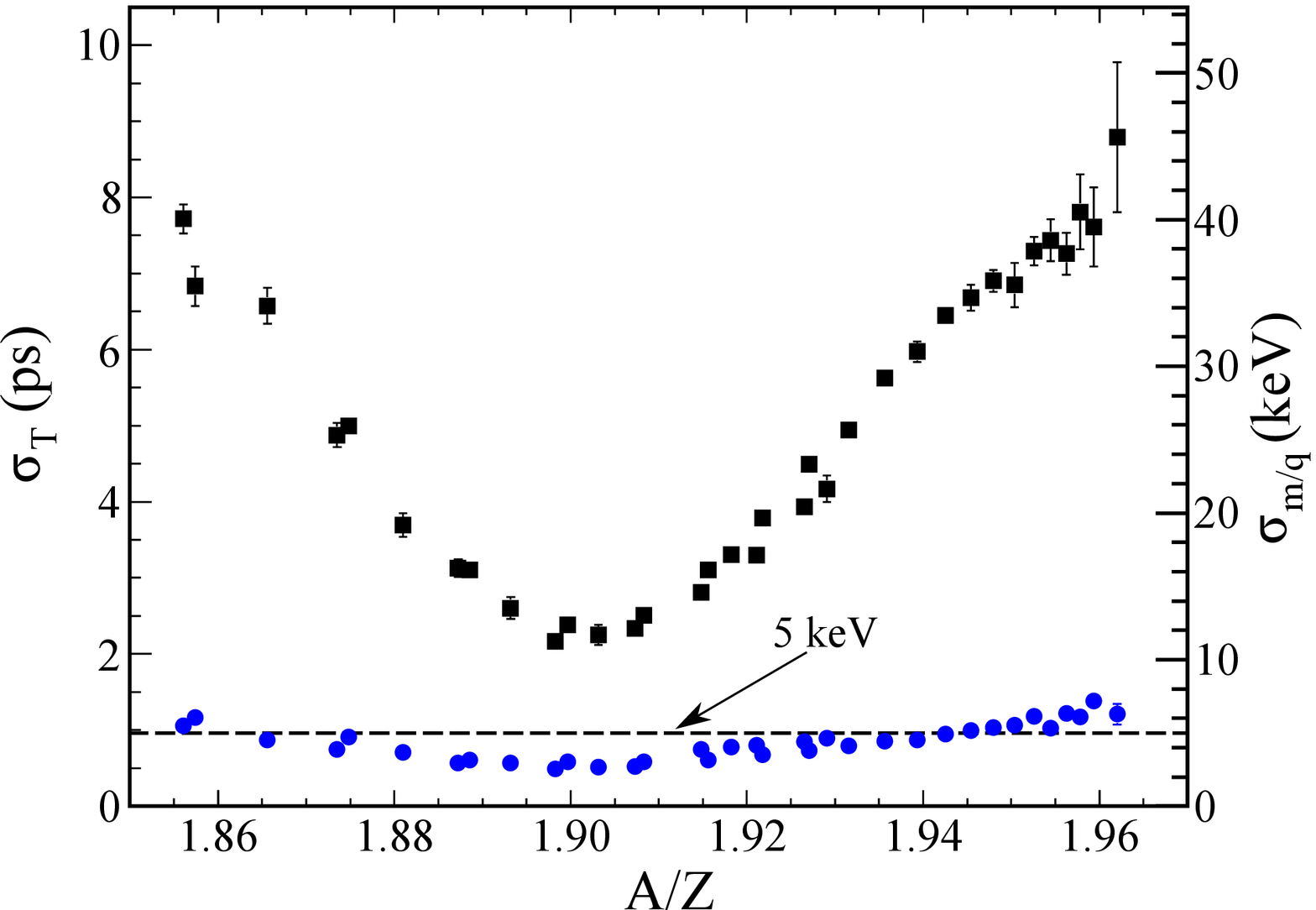}
	\caption{Standard deviations of the TOF peaks (left scale) derived from the original revolution time spectrum (black filled squares),
	and from the newly-constructed spectrum (blue circles). 
	The corresponding absolute accuracies of mass-to-charge ratios are given on the right scale.
	}
	\label{fig:st_m}
\end{figure}

The standard deviations of the $T$ peaks derived from the corresponding spectra are shown in Fig.~\ref{fig:st_m}. 
The standard deviations, $\sigma_T$, of the $T$ peaks in the original $T_{\rm exp}$ spectrum have a parabolic dependence versus $m/q$.
$\sigma_T$ approaches minimum at approximately 2 ps only for a limited number of nuclides (isochronicity window). 
In the spectrum obtained with the new $B\rho$-IMS technique $\sigma_T=0.5$ ps is achieved in the isochronicity window, 
corresponding to the mass resolving power of $3.3\times 10^5$ (FWHM). 
At the edges of the spectrum, a mass resolving power of $1.3\times 10^5$ (FWHM) can be achieved, improved by a factor of about 8 compared to the conventional method.
We emphasize that this was done without reducing the $B\rho$-acceptance of either the ring or the transfer line.
The right scale in Fig.~\ref{fig:st_m} shows the corresponding absolute $m/q$ precision.
It is emphasized that the $m/q$ precision of 5~keV can be obtained for just a single stored ion.

Our new mass results are in excellent agreement with the recent LEBIT PT measurements
for $^{44g,m}$V~\cite{Puentes2020} and $^{56}$Cu~\cite{Valverde2018}, 
%$^{51}$Fe~\cite{Ong2018}
and have a comparable precision. 
We note that the LEBIT measurements were performed one species at a time
~\cite{Puentes2020,Valverde2018},
%different settings had to be employed at LEBIT~\cite{Puentes2020,Valverde2018},
in contrast to just a single setting for all results in this work.

The newly-determined mass value of $^{52g}$Co is in good agreement with our earlier value~\cite{Zhang2018,Xu2016},
while there is a deviation with respect to the JYFLTRAP PT result~\cite{Nesterenko2017}. 
With our mass for $^{52g}$Co the proton separation energy, which is of interest for rp-process~\cite{Ong2018}, is modified to $S_p(^{52}{\rm Co})=1452(6)$~keV . 

The relative precision of (5 - 6)$\times10^{-8}$ has been reached for $^{46}$Cr, $^{48}$Mn, $^{50}$Fe and $^{54}$Ni nuclides,
making them the heaviest $T_z = -1$ nuclei for which the masses are known with such high precision.
Especially, $^{46}$Cr, $^{50}$Fe and $^{54}$Ni are the heaviest $T_z =-1$ superallowed $0^+\to 0^+$ $\beta$-emitters
used to calculate the $V_{ud}$ element of the Cabibbo-Kobayashi-Maskawa quark mixing matrix,
which should be unitary in the Standard Model~\cite{Hardy2020}.
By using our new mass results and the masses of the corresponding $\beta$-decay daughter nuclides~\cite{Wang2021},
the $\beta$ decay energies $Q_{EC}$ have been obtained.
In combination with the existing experimental data on the half-lives and branching ratios~\cite{Hardy2020,Molina2015},
the corrected $Ft$ values were determined for $^{46}$Cr, $^{50}$Fe and $^{54}$Ni, see Table~\ref{tab:ftvalue}.
The uncertainties stemming from the $Q_{EC}$ values are now one order of magnitude smaller than those from other contributions, such as half-lives and branching ratios.
Although the present uncertainties are still much larger than for other superallowed $0^+\to 0^+$ $\beta$ decays~\cite{Hardy2020},
the new data are essential for
tests of the theoretical corrections employed in Refs.~\cite{Hardy2020,Towner2015} by comparing the $Ft$ values of the decays presented in Table 2 with those of their mirror superallowed decays.
% tests of the mirror corrections used in Refs.~\cite{Hardy2020,Towner2015}. 
Higher-precision measurements of the half-lives and especially the branching ratios are needed in order to satisfy the requirements for a stringent test of the Standard Model.
\begin{table}[hbt]
    \caption{The determined superallowed $Q_{EC}$ values,
    statistical rate function $f$ in comparison with the data from the latest survey~\cite{Hardy2020},
    and the corrected values $Ft$.}\label{tab:ftvalue}
    \begin{ruledtabular}
    \begin{tabular}{ccccc}
        Fermi transition	& $Q_{EC}$ (keV)	& $f$	& $f$ in~\cite{Hardy2020}	& $Ft$ (s) \\
        \hline
        $^{46}{\rm Cr} \to {\rm ^{46}V }$	& 7598.7(2.6)	& 10647(20)	& 10685(74)	& 3130(100)\\
        $^{50}{\rm Fe} \to {\rm ^{50}Mn}$	& 8151.0(2.9)	& 15067(39)	& 15060(60)	& 3120(71)\\
        $^{54}{\rm Ni} \to {\rm ^{54}Co}$	& 8724.7(2.7)	& 21042(35)	& 21137(57)	& 3063(50)\\
    \end{tabular}
    \end{ruledtabular}
\end{table}

In summary, dramatically improved isochronous mass spectrometry, the $B\rho$-IMS, has been pioneered at the experimental cooler-storage ring CSRe.
Owing to simultaneous measurement of the revolution time and velocity of every stored short-lived ion,
the sensitivity and precision of the mass measurements were significantly increased.
The time sequences from the two TOF detectors are unique for each ion.
Only a few tens of 
%Merely a few ten 
signals are sufficient for unambiguous ion identification.
This unparalleled property of the $B\rho$-IMS makes it--in principle--a background free technique.
The overall measurement time is shorter than 1~ms, thus indicating that all $\beta$-decaying nuclei can be studied without lifetime restrictions.
The high mass resolving power was achieved over the whole $B\rho$-acceptance of the storage ring,
meaning that a large range of $m/q$ values can be covered in a single machine setting.
An uncertainty band as small as $\approx5$~keV was obtained.
This is a remarkable achievement,
indicating that storage of a single short-lived ($T_{1/2}\gtrsim100~\mu$s) ion is now sufficient for its mass determination with $\approx5\cdot q$~keV precision.
The $B\rho$-IMS is thus the ideally suited technique for high-precision mass measurements of the most exotic nuclides,
which have the shortest half-lives and tiniest production yields.

The first experimental results obtained by using the $B\rho$-IMS are reported here.
Masses of $^{46}$Cr, $^{50}$Fe and $^{54}$Ni were measured with relative mass precision of (5 - 6)$\times10^{-8}$.
The achieved mass precision is comparable to 
%the one 
that
reported for short-lived nuclei by Penning trap spectrometers.
The present measurements are for the heaviest $T_z = -1$ nuclei for which the masses are known with such a high precision.
New data were used to improve $Ft$ values that may be utilized for testing the Standard Model.

The merits of the $B\rho$-IMS, namely high resolving power, ultimate sensitivity, short measurement time, broadband and background-free, make it the technique of choice for future storage ring mass spectrometry.
The $B\rho$-IMS will be implemented at the Spectrometer Ring (SRing)~\cite{WU2018} of the high-intensity Heavy-Ion Accelerator Facility (HIAF)~\cite{YANG2013,Liu2018}, which is under construction in China.
It will also be used within the ILIMA project~\cite{Litvinov2010,WALKER2013} at the future FAIR facility in Germany~\cite{Bosch-2013,Durante-2019},
where two TOF detectors will be installed in a straight section of the Collector Ring CR~\cite{DOLINSKII2008,DOLINSKII2007}.

\begin{acknowledgments}
	The authors thank the staff of the accelerator division of IMP for providing stable beam. Comments and suggestions of Phil M. Walker are greatly appreciated. 
	This work is supported in part by the National Key R$\&$D Program of China (Grant No. 2018YFA0404401), 
	the Strategic Priority Research Program of Chinese Academy of Sciences (Grant No. XDB34000000), 
	and the NSFC (Grants No 12135017, No. 12121005, No. 11961141004, No. 11905259, No. 11905261, No. 11975280). 
	Y.M.X. and C.Y.F. acknowledge the support from CAS ``Light of West China" Program. 
	%R.J.C. is supported by the International Postdoctoral Exchange Fellowship Program 2017 by the Office of China Postdoctoral Council (No. 60 Document of OCPC, 2017). 
	Y.A.L. and R.S.S. are supported by the European Research Council (ERC) under the EU Horizon 2020 research and innovation programme (ERC-CG 682841 ``ASTRUm"). 
	T.Y. and S.S. were supported in part by JSPS and NSFC under the ``Japan-China Scientific Cooperation Program".
\end{acknowledgments}

% \bibliography{Brho_refs}

\end{document}